\documentclass{elsart}

 \usepackage{subfigure}
 \usepackage{graphics}
 \usepackage{graphicx}
\usepackage{lscape} 
\usepackage{rotating}
\usepackage{amssymb}

\def\Journal#1#2#3#4{ {#1} {\bf #2} #3 (#4)}

\def\NPBP{{\em Nucl. Phys. B (Proc. Suppl.)}}

\def\JCAP{\em JCAP}

\def\NPA{{\em Nucl. Phys.} A}
\def\PLB{{\em Phys. Lett.} B}

\def\PRL{\em Phys. Rev. Lett.}

\def\PRC{{\em Phys. Rev.} C}

\def\PRP{{\em Phys. Rep.}}

\def\ARAA{{\em Annu. Rev. Astron. Astrophys.}}

\def\APJ{\em Astrophys. J.}

\def\RPP{\em Rep. Prog. Phys.}

\def\be{\begin{equation}}
\def\ee{\end{equation}}
\def\bea{\begin{eqnarray}}
\def\eea{\end{eqnarray}}
\def\beas{\begin{eqnarray*}}
\def\eeas{\end{eqnarray*}}

\newcommand{\zbb}{2\mbox{$\nu\beta\beta$-decay~}}

\newcommand{\moeh}{\mbox{$^{100}$Mo}}
\newcommand{\ndhf}{\mbox{$^{150}$Nd} }
\newcommand{\pmhf}{\mbox{$^{150}$Pm} }
\newcommand{\smhf}{\mbox{$^{150}$Sm} }
\newcommand{\cft}{\mbox{$^{14}$C }}
\newcommand{\kref}{\mbox{$^{85}$Kr }}
\newcommand{\artn}{\mbox{$^{39}$Ar }}

\newcommand{\gdhs}{\mbox{$^{160}$Gd}}

\newcommand{\cdhs}{\mbox{$^{116}$Cd}}

\newcommand{\seza}{\mbox{$^{82}$Se}}

\newcommand{\yhss}{\mbox{$^{176}$Yb}}

\newcommand{\bes}{\mbox{$^7$Be}}
\newcommand{\nel}{\mbox{$\nu_e$}}

\begin{document}

\begin{frontmatter}




\title{Real-time spectroscopy of solar pp neutrinos using \ndhf}

\author[dresden]{K. Zuber}
\address[dresden]{Institut f\"ur Kern- und Teilchenphysik, Technische Universit\"{a}t Dresden, Germany}

\author{}

\address{}

\begin{abstract}
The potential real-time spectroscopy of solar pp neutrinos using \ndhf as target is investigated. The threshold
of 196 keV would be the lowest of all solar neutrino experiments running so far. Experimental rates and parameters 
are discussed, about 580 SNU can be expected from pp-neutrinos and another 367 SNU from $^7$Be. 
Furthermore, it is investigated whether charged current reactions might cause a new background component
for future double beta decay experiments based on a large amount of \ndhf.
\end{abstract}

\end{frontmatter}

\section{Introduction}
\label{intro}
In the last decade neutrino physics made terrific progress 
by establishing a non-vanishing neutrino mass. This result
stems from various neutrino oscillation searches including 
reactors, accelerators, the atmosphere and the Sun, for a
recent review see \cite{gon08}.
The latter, namely the problem of missing solar neutrinos was, after the pioneering
observation of the Homestake chlorine-experiment \cite{cle98}, one of the longest
standing problems in particle astrophysics. Various astrophysical
and particle physics solutions were proposed, but the problem
was finally settled by the gallium experiments GALLEX and SAGE, 
Super-Kamiokande and the Sudbury Neutrino Observatory SNO to be due to neutrino
oscillations in matter. Independently, the large mixing angle solution
was singled out as the only solution with the KamLAND detector observing spectral distortions
in reactor antineutrinos events. Recently,
the first real time detection of sub-MeV solar neutrinos in form of the \bes~ 
line has been published by
Borexino \cite{arp07,bel11}. This has opened the window of real-time observations of 
sub-MeV solar neutrinos. \\
While the basic solution of the solar neutrino problem has been found,
there are still a lot of issues for astrophysics and particle physics to be 
explored. First of all the satisfying agreement of solar neutrino observations 
with helioseismological
measurements and the Standard Solar Model has recently be worsened by
improved 3D fitting of photospheric lines \cite{asp09}. The newly deduced elemental
abundances lead to a worse description of helioseismological observations. Thus, one of the
fundamental assumptions of stellar structure physics, the homogeneous distributions
of the elements in stars, is in question and the measurement of neutrinos from the CNO cycle
can give unique information on the abundance of these elements in the solar interior.
Also particle physics will benefit from future solar neutrino measurements. Besides
matter oscillations alternative scenarios like non-standard interactions (NSI),
Mass Varying Neutrinos (MaVaNs) and potential contributions from Lorentz- and CPT-violatioin
have been proposed as alternative solutions
to the solar neutrino problem \cite{fri04,far04,dia11}. However, all of them propose a different
survival probability for \nel~ as the function of energy. The effect is most prominent
in the transition between vacuum and matter oscillations, i.e. around 1-2 MeV. Thus, an accurate
pep solar neutrino measurement is also very important. No measurement exists yet
but it could be done by SNO+ and will help to reduce the uncertainties on the mixing angle $\theta_{12}$. 
Also the small flux of hep neutrinos extending to highest energies
has not been observed yet and would complete the picture of our understanding
of stellar energy production.
Last but not least, the all over fundamental pp-neutrino flux, directly coupled with 
solar luminosity has not been observed in real time. Its observation and monitoring
will shed some light on the dynamics of the solar interior and any other time dependent effect.
All together, real time measurement of the individual fluxes will also determine the ratios of the
various branches of the fusion chains and a real-time measurement of the full solar neutrino spectrum
is the ultimate information one can get. For a recent review see \cite{tur11}.\\
By far the largest flux is that of pp-neutrinos originating from the fundamental fusion of two protons into deuterium
within the dominant pp-chain. 
It is directly coupled to the solar luminosity and Standard Solar Models predict a flux of 6 $\times 10^{10}$ neutrinos 
per cm$^{-2}s^{-1}$ with an error of about 0.5\% \cite{pen08}. Unfortunately for experiments
this is also the flux with the lowest energy terminating at neutrino energies of 423 keV \cite{bah97}. 
The only existing measurements are based on radiochemical methods
using $^{71}$Ga namely GALLEX and the follow up GNO \cite{alt05} as well as SAGE \cite{abd09}. 
For detection via neutrino-electron scattering this results
in a maximal energy of the electrons of about 233 keV. Potential real-time pp 
measurements via this process using existing large scale scintillators suffer
from backgrounds like \cft , \artn and \kref. 
Especially dangerous is the $\beta$-decay of \cft with a half-life of 5730 years and a
Q-value of 156 keV. Thus, for a long time other options in form of nuclear transitions were explored who
would allow radiochemical detection and spectroscopy of low energy solar neutrinos in real-time.
Among them are double beta emitters and long-living isotopes, the most promising
one for the latter is $^{115}$In with a threshold of 114 keV \cite{rag76,pfe78} currently under investigation
for the LENS experiment \cite{gri07}. From double beta candidates \moeh (threshold 168 keV),
\seza (threshold 173 keV), \gdhs (threshold 244 keV) \yhss (threshold 301 keV) and \cdhs (threshold at 464 keV, 
just above the pp-flux) were proposed 
\cite{rag97,eji00,zub03}.\\
In this paper a new candidate is explored for real-time spectroscopy of pp-neutrinos, namely \ndhf ,
a system also studied and used for double beta decay searches.


  


\section{The case for \ndhf and estimated rates}
A well known isotope of interest for double beta decay searches is \ndhf . 
It double beta decays via \pmhf into \smhf. Astonishingly according to \cite{toi98} no excited
state of \pmhf is known and even the ground state quantum numbers have some uncertainty, very like being
a 1$^-$-state. However, recently within charge exchange reactions studies
a single $1^+$ has been identified in \pmhf about 0.11 MeV  $\pm$ 10\% above the ground state with a 
Gamow-Teller strength of $B_{GT} = 0.13 \pm 0.02$
 \cite{gue11}. Furthermore, they suggest for the ground state a $2^-$ assignment of quantum numbers
 compared to the $1^-$ recommended in  \cite{toi98}.
The important point is that the newly discovered $1^+$-state will allow the detection of solar neutrinos
with an energy  threshold of 196 keV by neutrino capture on \ndhf, given that the fact the $Q_{EC}$ for the
electron capture of \pmhf is 86 keV \cite{aud03}.\\
To estimate a rate only solar pp-neutrinos and $^7$Be neutrinos are considered. 
The flux used for $^7$Be is from the latest Borexino measurement and given as 4.87 $\times 10^{9}$ cm$^{-2}s^{-1}$
\cite{bel11}. 
The pp-flux above the threshold of detection would be about 77\% of the total pp-flux. This has to be folded with
the survival probability of $\nu_e$ coming from the Sun, which is according to latest survival probability fits about
54-55\%.  Hence, the pp-flux considered for detection is  2.5 $\times 10^{10}$ cm$^{-2}s^{-1}$.
The absorption cross section can be written as \cite{boe92}
\begin{equation}
\sigma = 1.67 \times 10^{-45}  \langle p_e E_e F(Z,E_e) \rangle \mbox{cm$^2$}
\end{equation}
with $p_e,E_e$ as the momentum and energy of the electron in units of electron mass and $F(Z,E_e)$ 
as the Fermi function. The bracket takes into account a spectral averaging for the pp-neutrinos.
For the relativistic Fermi-functions the equations given in \cite{doi85} is used. 
Given the above values a rate of $\sigma \times \phi$ = 353 SNU for the 862 keV \bes~line (another about
14 SNU might come from the 384 keV line) can be determined, with the solar neutrino unit SNU being
 $10^{-36}$ captures per
target atom per second. Assuming a constant survival probability in the pp-region and a more or less 
constant flux in this region another $\sigma \times \phi \approx$ 580 SNU from pp-neutrinos can be
estimated. Thus, about 1000 kg of \ndhf enriched to 90 \% would result in roughly 104 events/yr. 
Even for a very small detector from the solar neutrino point of view, there is already a significant rate.


\section{Discussion of the expected signal }
\begin{figure}
\begin{center}
  \includegraphics[width=12cm,height=8cm]{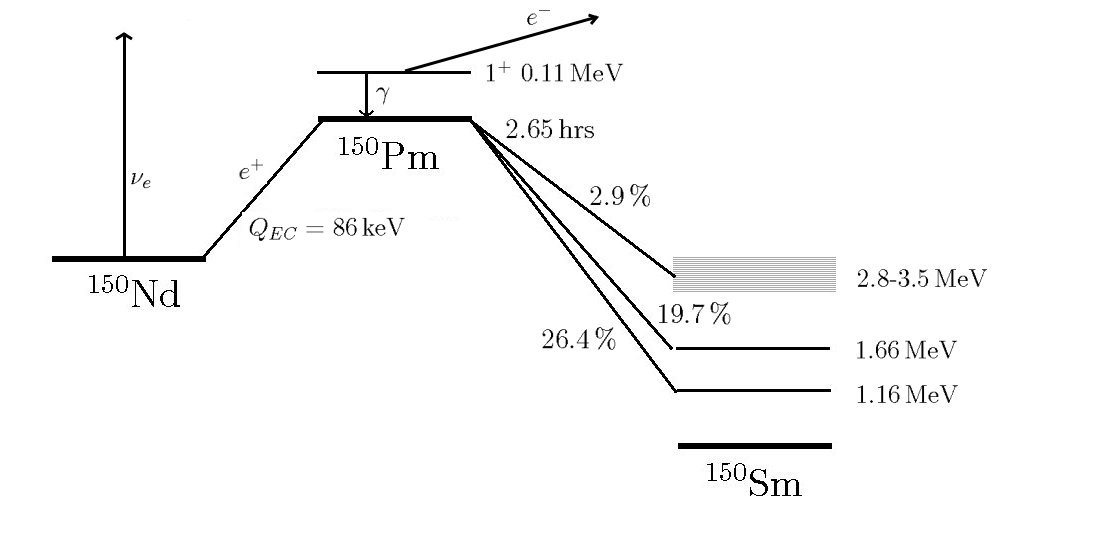}
\caption{\small Schematic drawing of the coincidence signal used for low energy solar neutrino
spectroscopy on \ndhf.}
\label{pic:coincidence}
\end{center}
\end{figure}

Like in several other radiochemical approaches discussed the neutrino capture will result in 
a coincidence, which is very convenient. The signal is shown in Fig.~\ref{pic:coincidence}.
First of all there will be an electron within an energy 
range of 0-227 keV together with a 110 keV de-excitation gamma which will be
followed by the $\beta$-decay of \pmhf. Whether this time coincidence can be used depends crucially
on the type of detector used as the half-life of \pmhf is 2.68 hours \cite{toi98}.  Otherwise one has to rely on one of the signal parts.
In liquid scintillator based approaches the coincidence search is very unlikely,
thus either the first part or the \pmhf decay can be used. \\

The produced radioisotope \pmhf has a complex decay scheme and will preferentially decay 
into excited states of  \smhf
emitting further characteristic gamma rays, in 68\% of all cases one of 333.92 keV.
Other observable gamma lines with more than 10\% emission probability are
at 1324.1 keV (17.5\%), 1165.73 keV (15.8 \%) and 831.85 keV (11.9\%).
In the following it is
assumed that the \pmhf decay will be used due to its higher energy release.
The dominant decay mode will be in 26.4\% of the cases into a 1$^-$-state at 1165.73 keV.
Furthermore 19.7 \% will decay into a (2$^{-}$)-state at 1658.41 keV and 17.8\% in a
(2$^{-}$)-state at 2070 keV with the accompanied electron, accounting for 64\% of the total 
decays. It should be mentioned in 12.4\% of the decays a total energy in form of gammas is emitted
within an energy range of 2100  $\leq E_\gamma \leq$ 2680 keV and in 2.91 \% of the cases
with more than 2800 keV total gamma energy. \\

The decay scheme of \pmhf is complex and the de-excitation energy
is released in several gammas. All transitions are allowed transition,
independent of the uncertain spin-parity assignment of the \pmhf ground state (except that
in the case of a $2^-$ ground state this would be purely Gamow-Teller type), thus
the energy spectrum of the electrons can be well described by the known form. 

\section{Experimental considerations}

As mentioned before the signal consists of a low energy electron (in case of $^7$Be neutrino 
capture it is in 90\% a monoenergetic 666 keV and in 10\% a 188 keV electron) 
in coincidence with a 110 keV gamma possibly
in a long time coincidence with the beta decay of \pmhf, resulting in a second electron and associated gammas.
As it is questionable whether any experiment will be designed especially for this purpose it might be
worthwhile to explore what the next generation of large scale experiments for double beta decay
based on \ndhf can do.\\

 Currently two kind of approaches are considered, Nd-foils spanned into TPCs
(the experiments DCBA and SuperNEMO) or Nd-loaded scintillators (SNO+).
Consider the case of SNO+ first, which is supposed to run in a first phase with 1000 tons of liquid scintillator 
with a 0.1\% natural Nd loading (total mass of 760 kg) and in a later stage with enriched \ndhf. 
As the produced gammas and electrons won't be resolved spatially the signal will be two energy depositions, 
the first part being the neutrino capture and thus
an energy deposition in the range of 110-337 keV (for pp-neutrinos) and 298 and 776 keV (for the $^7$Be lines)
respectively. The second part would be the \pmhf decay with a Q-value of 3454 keV. Due to the relative long
life-time of \pmhf of 2.68 hrs it is unlikely that a coincidence search can be used due to potential convection
and an overwhelming \zbb background. The known half-lives
of \ndhf ground state and first excited 0$^+$-state of this decay mode \cite{arg09} will lead to a rate of 0.5 Bq. 
In addition, this will also
swamp the low energy signal due to neutrino capture. Thus the only potential hope could be to search for the
high energy part of the \pmhf decay leading to events beyond 3 MeV. Relying solely on that the clear solar signal is
gone. Various other contributions will produce events in this energy range like $^{208}$Tl contaminations,
electrons from neutrino-electron scattering
produced by $^8$B solar neutrinos and direct production of \pmhf. For example \pmhf can be produced by (p,n) 
reactions on \ndhf.
However, the in-situ production by protons will be small as protons first of all have to be
created inside the detector by nuclear reactions and after that  the (p,n) reaction on \ndhf has to occur,
recently measured cross sections are about 30 mb for 10 MeV protons  \cite{rez11}.\\

In the described first phase of SNO+  the used amount of \ndhf is anyhow too small for 
any detection, the described problems for detection remain the same even for an enriched phase.
Here the background due to \zbb is even
orders of magnitude higher.\\

The second approach would be thin foils spanned within TPCs. Here the electrons could be tracked, even though
at these low energies the energy measurements might be disturbed by energy losses in the foil itself and
the observation crucially depends on the threshold used for electron detection.
An advantage will be that the coincidence of the capture signal and the \pmhf decay can be used, resulting in two
electron tracks originating from the same point of a foil within a few hours. Combined with the given energy
constraints on the electrons and the detection of gammas in the TPC a clear signal should be observed.
The disadvantage of this approach are the space requirements because the foils must be very thin to 
allow the electrons to escape. Thus, at the moment it seems unrealistic to build a ton scale experiment based on enriched \ndhf using
 foils and hence the reaction rate will be too low for solar neutrino spectroscopy.\\
 
To sum it up, unfortunately planned large scale double beta decay experiments using \ndhf might be not suitable for 
low energy solar neutrino detection. Other detector concepts based on \ndhf have to be developed perhaps a 
highly granulated Nd-loaded scintillator could be an option.\\  

However, a relatively high rate of low energy solar neutrino captures could cause a severe background 
for double beta searches on \ndhf, especially if the coincidence described before cannot be used. The neutrinoless
double beta peak is expect at 3371 keV \cite{kol10}. While this perhaps is not an 
issue for experiments using enriched Nd-foils due to their relatively low mass,
it could cause trouble for calorimetric approaches like Nd-loaded liquid scintillators. 
Here the different decay channels cannot be resolved and only the sum energy of the transition is measured.
As the Q-value of the \pmhf decay is 3454 keV there is a significant overlay with the Nd double beta peak region.
However with the above given estimate a detector with 500 kg of \ndhf enriched to 90\%, which is a potential
scenario for SNO+ phase 2, there would be only 52 events per year in total. With the given branching ratios
of the \pmhf decay the number of events in a region above 3000 keV up to 3454 keV will be much less than 
one event per year.


\section{Summary}
\label{conclusion}
Real time solar neutrino spectroscopy still offers a lot of information for partice and astrophysics.
With the recent discovery of an excited $1^+$ state in \pmhf in charge-exchange reactions 
a new opportunity for low energy real time measurements of low energy solar neutrinos using \ndhf has
been opened. Rates were estimated and revealed that even for a relatively small detector
a significant numbers of events can be achieved. The solar neutrino capture will not cause
a major worry for current or planned double beta decay experiments.

\section*{Acknowledgement}
The author would like to thank R. Dvornicky for discussions and help with the Fermi-function
and O. Chvorets for valuable discussions. The help of B. Lehnert is also acknowledged.

\newpage

\begin{thebibliography}{00}
\bibitem{gon08} Gonzalez-Garcia~M~C and Maltoni~M \Journal{\PRP}{460}{1}{2008}
\bibitem{cle98} Cleveland B T \etal ., \Journal{\APJ}{496}{505}{1998}
\bibitem{arp07} Arpesella~C et al., \Journal{\PRL}{101}{091302}{2007}
\bibitem{bel11} Bellini~G et al., Preprint arXiv:1104.1816
\bibitem{asp09} Asplund~M \etal \Journal{\ARAA}{47}{481}{2009}
\bibitem{fri04} Altmann~A, Lunardini~C and Pena-Garay~C, \Journal{\PLB}{594}{344}{2004}
\bibitem{far04} Fardon~R, Nelson~A~E and Weiner~N, \Journal{\JCAP}{0410}{005}{2004},
Preprint astro-ph/0309800
\bibitem{dia11} Diaz~J~S and Kostelecky~A, Preprint arXiv:1108.1799
\bibitem{tur11} Turck-Chieze~S., Couvidat~S, \Journal{\RPP}{74}{086901}{2011}
\bibitem{pen08} Pena-Garay~C, Serenelli~A, Preprint arXiv:0811.2424 
\bibitem{bah97} Bahcall~J~N, \Journal{\PRC}{56}{3391}{1997}
\bibitem{alt05} Altmann~M et al., \Journal{\PLB}{616}{174}{2005}
\bibitem{abd09} Abdurashitov~J~N et al., \Journal{\PRC}{80}{015807}{2009}
\bibitem{rag76} Raghavan~R, \Journal{\PRL}{37}{259}{1976}
\bibitem{pfe78} Pfeiffer~L et al., \Journal{\PRL}{41}{63}{1978}
\bibitem{rag97} Raghavan~R, \Journal{\PRL}{78}{3618}{1997}
\bibitem{gri07} Grieb~C, \Journal{\NPBP}{168}{122}{2007}
\bibitem{eji00} Ejiri~H et al., \Journal{\PRL}{85}{2917}{2000}
\bibitem{zub03} Zuber~K, \Journal{\PLB}{571}{148}{2003}
\bibitem{toi98}The 8th edition of the Table of Isotopes, Richard B. Firestone, Virginia S. Shirley, Coral M. Baglin, S.Y. Frank Chu, and Jean Zipkin. John Wiley \& Sons Inc., 1996.
\bibitem{gue11}  Guess~C~J et al., \Journal{\PRC}{83}{064318}{2011}
\bibitem{aud03} Audi~G, Wapstra~A~H and Thibault~C,  \Journal{\NPA}{729}{337}{2003}
\bibitem{boe92} Boehm~F, Vogel~P., {\it Physics of massive neutrinos}, Cambridge Univ. Press, 1992 
\bibitem{doi85} Doi~M, Kotani~T and Takasugi~E, \Journal{PTPS}{83}{1}{1985} 
\bibitem{arg09} Argydiades~J et al., \Journal{\PRC}{80}{032501}{2009}
\bibitem{rez11} Lebeda~O et al., submitted
\bibitem{kol10} Kolhinen~V~S et al.,  \Journal{\PRC}{82}{022501}{2010}





\end{thebibliography}
\end{document}